\documentclass[twocolumn]{svjour3}

\usepackage{amsmath}
\usepackage{amssymb}
\usepackage{epsfig}
\usepackage{color}

\graphicspath{{.}{./EPS/}}

\newcommand{\braket}[2]{\langle{#1}|{#2}\rangle}
\newcommand* {\bra}[1]{\ensuremath{\langle {#1} |}}
\newcommand* {\ket}[1]{\ensuremath{| {#1} \rangle}}

\begin{document}
\title{
Unambiguous atomic Bell measurement assisted by multiphoton states
}

\author{Juan Mauricio Torres \thanks{\email{mauricio.torres@physik.tu-darmstadt.de}}%
\and J\'ozsef Zsolt Bern\'ad 
\and Gernot Alber 
}\institute{Institut f\"{u}r Angewandte Physik, Technische Universit\"{a}t Darmstadt, D-64289, Germany}

\date{\today}
\maketitle
\begin{abstract}
We propose and theoretically investigate an unambiguous
Bell measurement of atomic qubits assisted by multiphoton states.
The atoms interact resonantly with the electromagnetic field inside two spatially 
separated optical cavities in a Ramsey-type
interaction sequence. 
The qubit states are postselected by measuring the photonic states inside the resonators. 
We show that if one is able to project the photonic field
onto two coherent states on opposite sites of phase space, 
an unambiguous Bell measurement can be implemented.
Thus our proposal may provide a core element for future components 
of quantum information technology  such as a quantum repeater 
based on coherent multiphoton states, atomic qubits and  matter-field interaction. 
\end{abstract}

\section{Introduction}
\label{Intro}

Establishing well-controlled entanglement between spatially separated quantum systems is essential for 
quantum communication \cite{Briegel98,Duer99}. At its core a quantum repeater employs entanglement which is generated and distributed 
among intermediary nodes positioned not too distant from each other. Entanglement purification \cite{Bennett,Deutsch} enables the distillation of a high-fidelity state from a 
large number of low-fidelity entangled pairs and with the help of entanglement swapping procedures \cite{Zuk}
the two end points of a repeater are entangled. 
There are many different implementation proposals for quantum repeaters, utilizing completely different systems and entanglement distribution protocols \cite{Sangouard}.    
A promising approach towards these schemes is to require some compatibility with existing optical communication networks. The proposal of 
van Loock et al. \cite{vanLoock1,vanLoock2,vanLoock3,vanLoock4} is such an approach where the repeater scheme employs coherent multiphoton states. These proposals
assume dispersive interaction between the atomic qubits and the single-mode of the radiation field. This
imposes limitations on the photonic postselection. 
It was shown that these limitations can be overcome in the case of resonant atom-field interactions \cite{Bernad1,Bernad2}
and it was demonstrated
for one building block of a repeater, namely the entanglement generation between spatially separated and neighbouring nodes. A natural extension of this approach is
to propose resonant atom-field interaction based schemes also for the other building blocks. 
In the case of entanglement swapping a complete atomic Bell measurement is required.

  Bell measurements play a central role also in en\-tan\-gle\-ment-as\-sisted quantum 
teleportation \cite{Bennett93} and in superdense coding \cite{Bennett92}. 
In the case of photonic qubits theoretical proposals \cite{Knill,Pittman,Munro} have been made and 
experimental realizations have already been carried out \cite{Kim,Schuck}. 
However, for atomic qubits there are still experimental difficulties which restrain implementations of complete Bell measurements
where projections onto the four Bell states can be accomplished. 
There exist experimental proposals that rely on the application of a
controlled NOT gate \cite{Pellizzari,Lloyd}. These proposals have the drawback that 
experimental implementations of two-qubit gates  have still complications to attain
high fidelity \cite{Schmidt-Kaler,Isenhower,Noelleke}. 
This implies that the fidelity of the generated Bell states 
is also affected \cite{Isenhower}. 
A proposal focusing specifically on a non-invasive atomic Bell measurement with high fidelity is still missing.

In previous work we have introduced a protocol to project onto one Bell state with
high fidelity \cite{Torres2014} based on atomic qubits which interact sequentially 
with coherent field states prepared in 
two cavities. The field states emerging after the interactions are postselected by balanced 
homodyne photodetection. In this paper we expand our previous work
to accomplish the projection onto all four Bell states provided the protocol is successful.
Thus we introduce an unambiguous Bell measurement of 
two atomic qubits with the help of coherent multiphoton field states. 
We demonstrate that the possibility of implementing field projections onto two coherent states on 
opposite sites of phase space implies the possibility to realize an unambiguous 
Bell measurement. Our protocol has a finite probability of error depending on the initial states
of the atoms. This is due to the imperfect overlap of the field contributions with  
coherent states. Nevertheless, it is an unambiguous protocol as there are four successful
events that lead to postselection of four different Bell states. 
The scheme is based on basic properties of the 
two-atom Tavis-Cummings model \cite{Tavis} and on resonant matter-field interactions which are already under 
experimental investigation \cite{Casabone1,Casabone2,Reimann,Nussmann2005}.  These considerations make our scheme compatible with a quantum repeater or a quantum relay 
based on coherent multiphoton states, atomic qubits and resonant matter-field interaction.  
Our proposal demonstrates that scenarios involving the two-atom Tavis-Cummings model are rich enough to enable future Bell measurement implementations.

The paper is organized as follows. In Sec. \ref{Model} we introduce the
theoretical model and  analyse the solutions of the field state with the aid of the Wigner function
in phase space.
Furthermore, we provide approximate solutions of the global 
time dependent  state vector that facilitate the  analysis of the system.
In Sec. \ref{Bell} we present a scheme to perform an unambiguous Bell
measurement provided one is able to project a single mode photonic field onto
coherent states. 
In Sec. \ref{Discussion} we provide a numerical analysis of the fidelity of the
projected Bell states and discuss general features of the protocol.
Details of our calculations are presented in  Appendices \ref{appendix} and 
\ref{appendixoverlap}.

\section{Theoretical model}
\label{Model}
\subsection{Basic equations}
In this section we recapitulate basic features of the two-atom Tavis-Cummings model 
\cite{Tavis}.
This model has been considered previously to study the dynamics of entanglement 
\cite{Torres2014,Jarvis,Rodrigues,Kim2002,Tessier}.
The model describes the interaction between two atoms $A$ and $B$ and
a single mode of the radiation field with  frequency $\omega$. The two identical atoms 
have ground states $\ket{0}_i$ and excited states 
$\ket{1}_i$ ($i \in\{A,B\}$) separated by an energy difference of $\hbar \omega$. 
In the dipole and rotating-wave approximation the Hamiltonian in the interaction picture 
is given by
\begin{align}
  \hat{H}&= 
\hbar g\sum_{i=A,B}\left( 
\hat{\sigma}^+_i\hat{a}+ 
  \hat{\sigma}^-_i\hat{a}^\dagger\right)
\label{Hamilton}
\end{align}
where 
$\hat{\sigma}^+_i=\ket{1}\bra{0}_i$ and $\hat{\sigma}^-_i=\ket{0}\bra{1}_i$
are the atomic raising and lowering operators
 ($i \in\{A,B\}$), 
 and $\hat{a}$ ($\hat{a}^\dagger$) is the annihilation (creation) operator  of the single mode 
 field. The coupling between the atoms and the  field is characterized by
 the vacuum Rabi frequency $2g$.

The time evolution of the system can be evaluated for an initial pure state as 
\begin{equation}
\ket{\Psi_t}=e^{-i\hat Ht/\hbar}\ket{\Psi_0}.
\label{EPsi}
\end{equation}
We are interested in the case where the atoms and the cavity are assumed 
to be prepared in the product state
\begin{equation}
  \ket{\Psi_0}=
  \Big(c^-\ket{\Psi^-}+
  c^+\ket{\Psi^+}+d_\phi^-\ket{\Phi^-_\phi}+
  d_\phi^+\ket{\Phi^+_\phi}
  \Big)
  \ket{\alpha},
  \label{initial}
\end{equation}
with the radiation field considered initially in a coherent state \cite{Glauber,Perelomov}
\begin{align}
  \ket{\alpha}=\sum_{n=0}^\infty 
  e^{-\frac{|\alpha|^2}{2}}
 \frac{\alpha^n}{\sqrt{n!}}
 \ket n,
  \quad\alpha=\sqrt{\overline n}\,e^{i\phi},
  \label{coherentstate}
\end{align}
with mean photon number $\bar n$ and photon-number states $\ket{n}$. 
The parameters $c^\pm$ and $d_\phi^\pm$ are the initial probability amplitudes of the
orthonormal  Bell states
\begin{align}
  \ket{\Psi^\pm}&=\tfrac{1}{\sqrt2}\left(\ket{0,1}\pm\ket{1,0}\right),
  \nonumber\\
  \ket{\Phi^\pm_\phi}&=\tfrac{1}{\sqrt2}\left(
  e^{-i\phi}\ket{0,0}\pm e^{i\phi}\ket{1,1}
  \right),
  \label{bellstates}
\end{align}
with the atomic states
$\ket{i,j}=\ket{i}_A\ket j_B$ ($i,j \in \{0,1\}$). We have chosen an atomic orthonormal 
basis containing the states $\ket{\Psi^\pm}$ as
the state $\ket{\Psi^-}\ket n$ is an invariant state of the system. This is explained in  Appendix \ref{appendix} where
we present the full solution of the temporal state vector. The other two Bell states $\ket{\Phi_\phi^\pm}$ depend
on the initial phase $e^{i\phi}$ of the coherent state. 
They appear naturally in the Tavis-Cummings model 
due to the exchange of excitations between atoms and cavity, and are involved in 
an approximate solution of the state vector that facilitate the analysis of the dynamics. 
Before showing the detailed form of our solution, 
let us give an overview of the dynamical features  that impose  relevant time scales in the system.

\subsection{Collapse and revival phenomena}
The collapse and revival phenomena
of the  Jaynes{\-}-Cum\-mings model and of the two-atom Tavis-Cummings model 
play an essential role in the quantum information protocols presented in
Refs. \cite{Bernad1,Torres2014,Jarvis}. 
This behaviour was first found in the time dependent atomic population in the
Jaynes-Cummings model \cite{Eberly1980}  when the field is initially prepared
in a coherent state: The populations display Rabi oscillations
that cease after a collapse time $t_c$ and appear again at a revival time $t_r$.
In the case of the two atom Tavis-Cummings model,
the collapse and revival time of the Rabi oscillations 
are given by
\begin{align}
  t_r=\frac{\pi}{g}\sqrt{4\bar n+2}, \quad 
  t_c=\frac{1}{\sqrt2 g}.
  \label{revcoltime}
\end{align}
These time scales have been previously introduced and can be found, for instance, in
Refs. \cite{Jarvis,Eberly1980}.
As they play an essential role in the dynamics of the system
it is convenient to introduce the rescaled time
\begin{align}
  \tau =t/t_r=tg/\pi\sqrt{4\bar n+2}.
  \label{tau}
\end{align}

\begin{figure}[h]
\begin{center}
\includegraphics[width=0.43\textwidth]{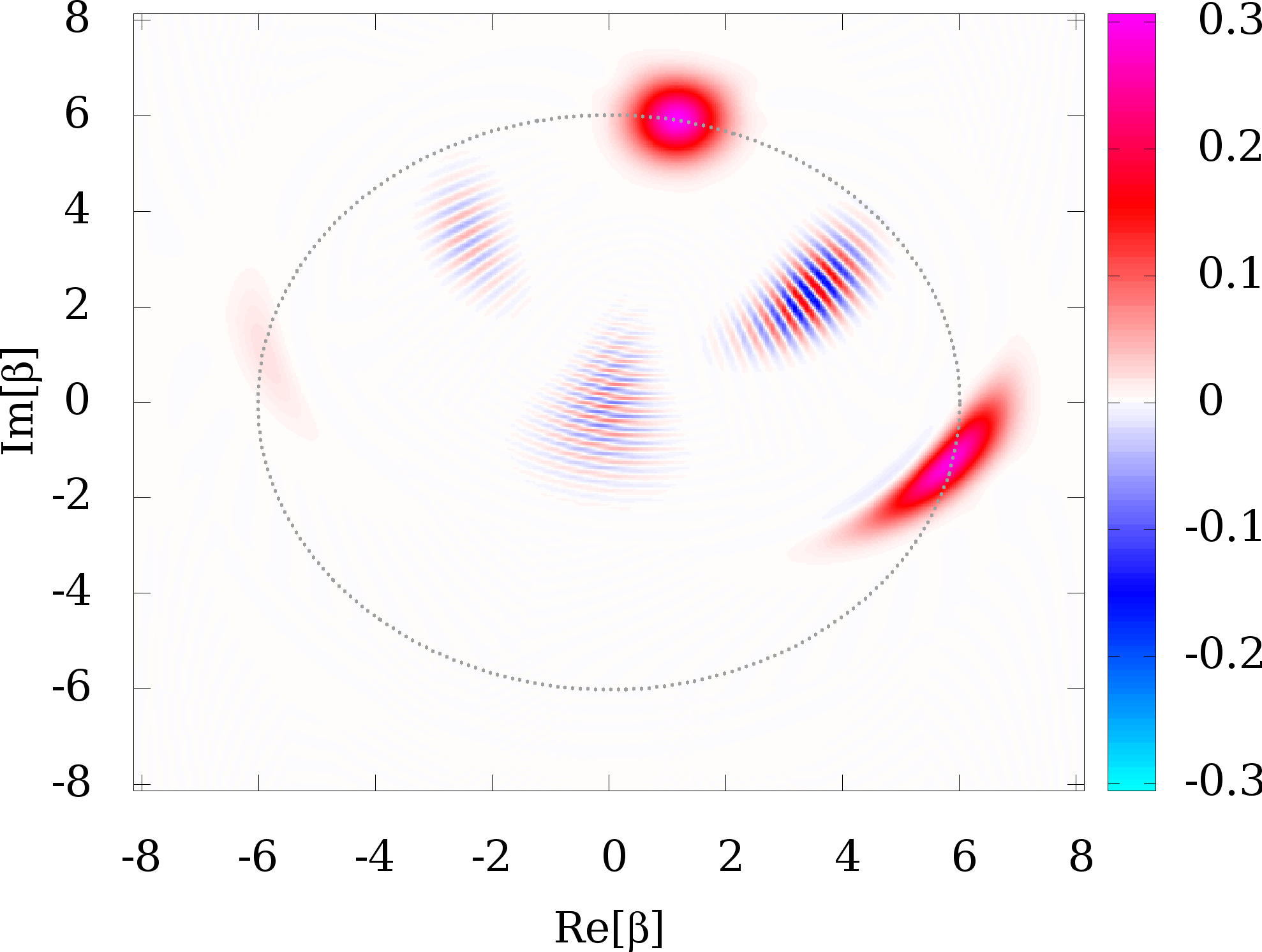}
\includegraphics[width=0.43\textwidth]{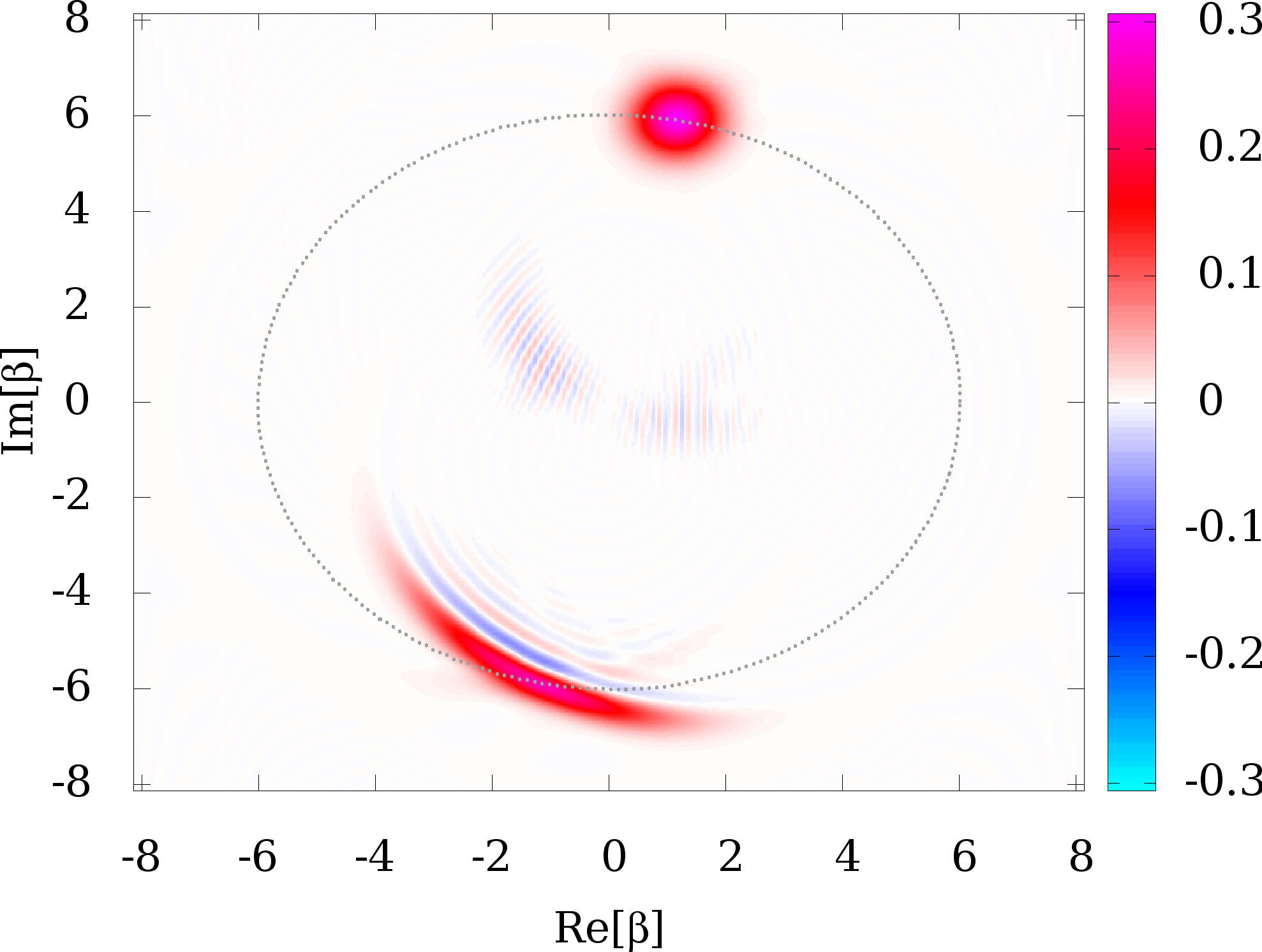}
\end{center}
\caption{\label{wigfig}
Wigner function of the cavity field after the interaction with two two-level atoms at times $\tau=1/4$ (top)
and $\tau=1/2$ (bottom) with the rescaled time of Eq. \eqref{tau}. 
The initial states of the atoms are defined by the parameters: 
  $c^-=0.5554$, 
  $c^+=0.3213+i0.5004$,    
  $d^-_\phi=-0.2053+i0.3726$,
  $d^+_\phi=0.1046+i0.3819$, and the parameter $\alpha=\sqrt{36.16}e^{i1.37}$ characterizes the
  initial coherent sate.}
\end{figure}

Let us explain these phenomena by visualizing the phase space of the radiation field 
with the aid of the Wigner function \cite{Schleich,Risken}
\begin{equation}
W_t(\beta,\beta^*) =
\frac{1}{\pi^2} \int {\rm Tr}\left\{\hat{\varrho}_t \, 
e^{\zeta \hat{a}^\dagger-\zeta^* \hat{a}}\right\}
 e^{\beta\zeta^*-\beta^*\zeta} d^2\zeta,
\label{Wignerf}
\end{equation}
with the complex numbers  $\beta$ and $\zeta$. The operator 
$  \hat{\varrho}_t ={\rm Tr}_{\rm atoms}\left\{\ket{\Psi_{t}}\bra{\Psi_{t}}\right\}$
is the density matrix of the field state obtained after taking partial trace over the atomic degrees of freedom 
from the full density matrix corresponding to the state vector in Eq. \eqref{EPsi}.
In Fig. \ref{wigfig} we show the Wigner function after interaction times  $\tau=1/4$ in the
top panel and $\tau=1/2$ in the bottom panel.
The circular shape corresponds to the initial coherent state $\ket{\alpha}$. This contribution
to the field remains stationary as long as there is an initial contribution of the state 
$\ket{\Psi^-}$. The reason is that $\ket{\Psi^-}\ket n$ is an invariant state of the system.
There are two other contributions to the field that rotate around the origin. In the top panel
of Fig. \ref{wigfig} it can be noticed that for an interaction time of $\tau/4$ they have
completed a quarter of cycle. At the bottom, the situation at interaction time $\tau/2$ is
shown where half a rotation has been completed.  
The interference fringes between the field contributions signify that there are coherent
superposition between these states of the field. The behaviour of the field state in phase space
explain the phenomena: Rabi oscillations cease (collapse) when the field contributions are well 
separated, e.g. at time $\tau/4$,
and revive when the field contributions overlap, e.g. at $\tau/2$ or the main revival at $\tau$ when
all the field constituents  coincide at the position of the initial coherent state.

\subsection{Approximation of the  state vector}
The full solution to the time dependent state vector of the two atoms Tavis-Cummings model
has already been presented in previous work, see for instance \cite{Kim2002,Torres2010}. Coherent
state approximations have also been considered  in the past 
\cite{Torres2014,Jarvis,Rodrigues,Gea-Banacloche}. 
In this context,
the eigenfrequencies of the Hamiltonian \eqref{Hamilton} that depend on the photonic number $n$
 are expanded in a first order Taylor series
around the mean photon number $\bar n$. However, the coherent state description is accurate only for
times well below the revival time.
In this work we go beyond the coherent state approximation by considering second order
contributions of the eigenfrequencies around $\bar n$. 
The details can be found
in the Appendix \ref{appendix} where it is shown that the time dependent 
state vector of the system can be approximated by
\begin{align}
  \ket{\Psi^{\rm A}_\tau}= &
  \frac{1}{N_\tau}\left(
  c^-\ket{\Psi^-}+
  d_\phi^-
  \ket{\Phi^-_{\phi}}
  \right)
  \ket\alpha+
  \nonumber\\
  &
  \frac{c^+-d^+_{\phi}}{2N_\tau}
  \left(
  \ket{\Psi^+}-\ket{\Phi^+_{\phi+2\pi\tau}}
  \right)
  \ket{\alpha_\tau^+}+
  \nonumber\\
  &
  \frac{c^++d^+_{\phi}}{2N_\tau}
  \left(
  \ket{\Psi^+}
  +\ket{\Phi^+_{\phi-2\pi\tau}}
  \right)
  \ket{\alpha_\tau^-},
  \label{Psit}
\end{align}
with the photonic states 
\begin{equation}
  \ket{\alpha_\tau^\pm}=
  \sum_{n=0}^\infty
  \frac{\alpha^ne^{-\frac{|\alpha|^2}{2}}}{\sqrt{n!}}
  e^{\pm i 2\pi\tau\left[\bar n+1+n-\frac{(n-\bar n)^2}{4\bar n +2} \right]}\ket n
  \label{PhotonicStates}
\end{equation}
and with the normalization factor  
\begin{align}
  N_\tau=\Big(1&+
  {\rm Re}[
(c^++d^+_\phi)^\ast
(c^+-d^+_\phi)
\braket{\alpha^-_\tau}{\alpha^+_\tau}
  ]\sin^2(2\pi\tau)
  \nonumber\\
  &+2
  {\rm Re}[d^-_\phi (d^{+}_\phi)^\ast]{\rm Im}
  [\braket{\alpha^-_\tau}{\alpha}]
  \sin( 2\pi\tau).
  \nonumber\\
  &+2{\rm Im}[(c^{+})^\ast d^-_\phi]{\rm Re}
  [\braket{\alpha^-_\tau}{\alpha}]
  \sin (2\pi\tau)\Big)^{1/2}.
  \label{normalization}
\end{align}
The quantity $\braket{\alpha^-_\tau}{\alpha}$
is evaluated in Appendix
\ref{appendixoverlap} and an approximate expression 
is given in Eq. \eqref{overlapepsilon}.

\begin{figure}
  \includegraphics[width=0.48\textwidth]{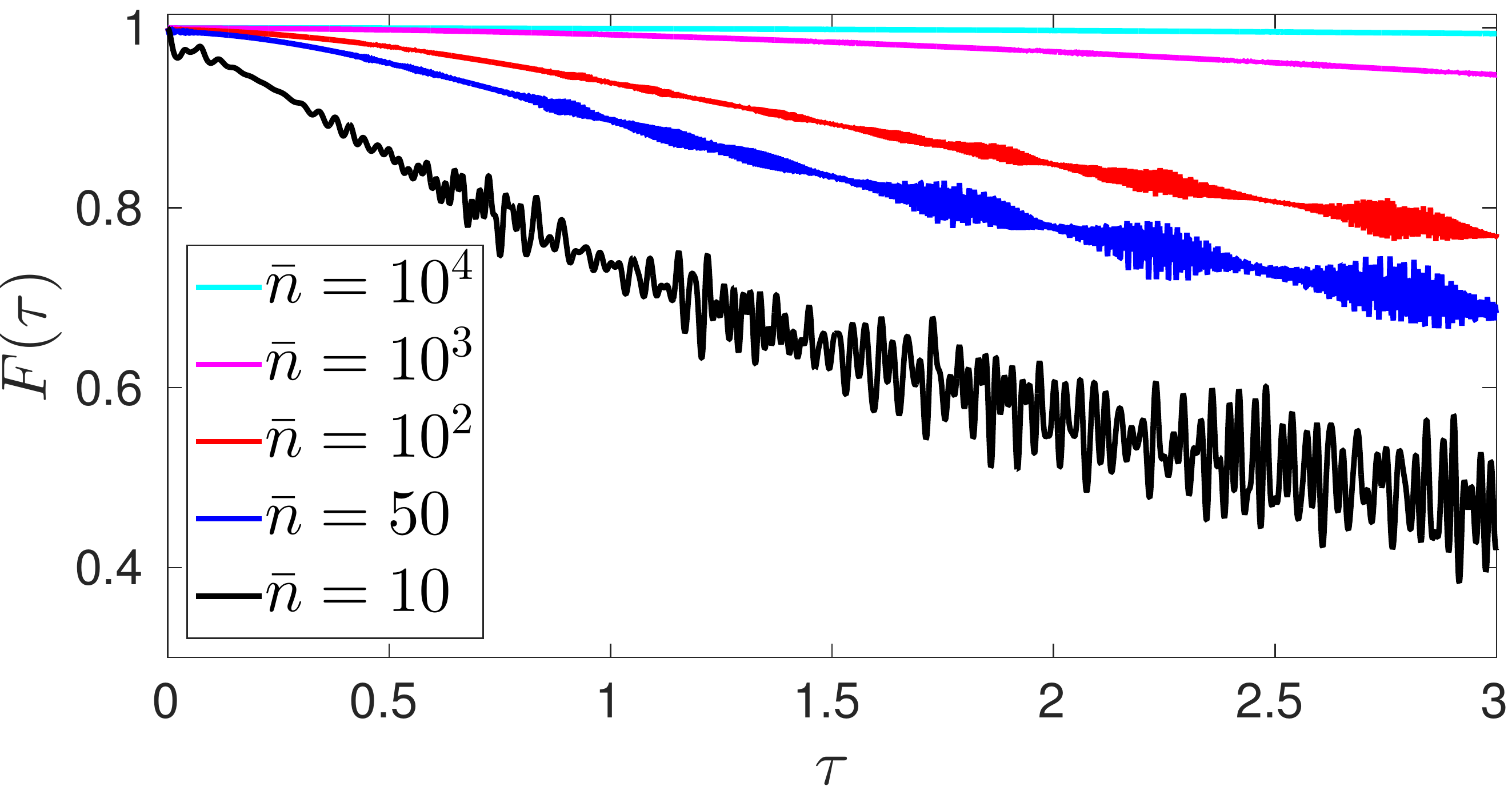}
  \caption{\label{totalfidelity}
  Fidelity of the total state of Eq. \eqref{Psit} with respect to the exact solution
  given by Eq. \eqref{psi} as a function of the time $\tau$ in Eq. \eqref{tau}
  scaled in terms of the revival time $t_r$: Five curves are presented for 
  different values of the mean photon number $\bar n$ as described in the legend.
  The rest of the parameters are the same as in Fig. \ref{wigfig}.
  }
\end{figure}

In order to test the validity of  Eq. \eqref{Psit} we have considered
the fidelity  
$F(\tau)=|\braket{\Psi^{\rm A}_\tau}{\Psi_{t_r\tau}}|^2$
of the approximated state vector with respect to the exact result given in Eq.
\eqref{psi}. 
In Fig. \ref{totalfidelity} we have plotted 
the results of numerical evaluations of the fidelity 
$F(\tau)$ for different values of the mean photon
number $\bar n$. 
It can be noticed that the validity of this approximation improves with
increasing mean photon number $\bar n$. 
In the Appendix \ref{appendix} it is 
discussed that 
our approximation is valid provided the condition 
$\tau\ll \sqrt{\bar n}/2\pi$ is fulfilled.

The form of the solution given in Eq. \eqref{Psit} allows a simple 
analysis of the dynamics. 
It is written in terms of an orthonormal atomic basis of Bell states
and is therefore suitable for the analysis of the atomic entanglement.
In particular, 
it is interesting to note that for an initial state without a contribution
of the state $\ket{\Phi^-_\phi}$, i.e. $d_\phi^-=0$, a 
photonic projection that discriminates the state  $\ket\alpha$ from the states
$\ket{\alpha_\tau^\pm}$ can postselect the atomic Bell state $\ket{\Psi^-}$.
In Ref. \cite{Torres2014} we studied this Bell state projection and found that
its implementation requires a flexible restriction for the interaction time:
it has to be below the revival time and above the collapse time 
given in Eq. \eqref{revcoltime}. 
In the following we concentrate in a more specific interaction time. We
analyze the dynamics at the specific interaction time $\tau=1/2$. 
This analysis  will allow us to introduce in Sec. \ref{Bell} 
a protocol to perform the four Bell state projections.

\vspace{.5cm}
\subsection{Basic dynamical features at scaled interaction time $\tau=1/2$}
There are two main reasons for studying in detail the case with   
 scaled interaction time $\tau=1/2$. The first one is that
the time dependent atomic states in Eq. \eqref{Psit} coincide,
i.e.
\begin{align}
\ket{\Phi_{\phi+\pi}^\pm}=\ket{\Phi_{\phi-\pi}^\pm}=-\ket{\Phi_\phi^\pm}.
  \label{property1}
\end{align}
The second reason is that the photonic states $\ket{\alpha_{1/2}^\pm}$
have completed  half a rotation in phase space and lie 
on the opposite site to the initial 
coherent state $\ket\alpha$
whereby overlapping with the coherent state $\ket{-\alpha}$. 
This means that at this interaction time
and for $|\alpha|\gg 1$, the initial photonic state 
$\ket\alpha$ can be approximately distinguished from the other two states $\ket{\alpha_{1/2}^\pm}$.
However, the states $\ket{\alpha_{1/2}^\pm}$ overlap significantly. 
This can be noticed in Fig. \ref{wigfig} where we have plotted the Wigner function.
The circular shape corresponds to the initial coherent state $\ket{\alpha}$, 
while the distorted ellipses on the opposite site of the phase space
correspond to the states $\ket{\alpha^\pm_{1/2}}$.
To distinguish these two components of the field it is convenient to conceive an experiment that is able
to project the field state onto the coherent states  $\ket{\pm\alpha}$.

In order to study the projection  onto the state $\ket{\pm \alpha}$, one has
to evaluate its overlap with the photonic states of the state vector in Eq. \eqref{Psit}.
First we consider the overlaps that can be neglected for large value $\bar n$,
namely
\begin{align} 
  \braket{\alpha}{-\alpha}= e^{-2\bar n},\quad
  \braket{\alpha}{\alpha_{1/2}^\pm}\propto e^{-\frac{2\pi^2}{4+\pi^2}\bar n}.
  \label{overlapsimple}
\end{align}
The explicit form of the overlap $\braket{\alpha}{\alpha_{1/2}^\pm}$
is given in Eq. \eqref{overlapapp} of the Appendix \ref{appendixoverlap} where
its approximation is also evaluated. 
The nonvanishing overlaps in the limit of large mean photon number are $\braket{\alpha}{\alpha}=1$
and
\begin{align}
  \braket{-\alpha}{\alpha_{1/2}^\pm}
  &\approx 
 \sqrt{\frac{2}{\sqrt{4+\pi^2}}}
 e^{\mp i\left(\frac{1}{2}\arctan\frac{\pi}{2}-(\bar n+1)\pi\right)}.
  \label{overlaptheta}
\end{align}
The expression  in Eq. \eqref{overlaptheta}
is also evaluated in detail in  Appendix \ref{appendixoverlap}. 
This overlap is real valued if the
mean photon number $\bar n=|\alpha|^2$ fulfills the relation
\begin{align}
  \bar n= m +\tfrac{1}{2\pi}\arctan\tfrac{\pi}{2},\quad {\rm with}\quad m\in \mathbb{N}.
  \label{ncondition}
\end{align}

If the condition in Eq. \eqref{ncondition} is fulfilled and if 
we suppose an initial atomic state with
no contribution from the state $\ket{\Phi^+_\phi}$, i.e. $d^+_\phi=0$, 
then it can be verified that a projection
onto the field state $\ket{-\alpha}$ postselects the atoms in the unnormalized atomic Bell state 
$\sqrt b c^+\ket{\Psi^+}$, with 
\begin{align}
b=2/\sqrt{4+\pi^2}. 
  \label{factor}
\end{align}
The success probability of this projection is given by $b|c^+|^2$ which
is proportional to the initial probability of this particular Bell state
$\ket{\Psi^+}$. The factor $b$ is the
result of our inability to project perfectly and simultaneously onto both field states 
$\ket{\alpha^\pm_{1/2}}$. 
In the next section we present a protocol that can perform
postselection of the four Bell states regardless of the initial state of the atoms.

\section{An unambiguous Bell measurement}
\label{Bell}
\begin{figure}
  \includegraphics[width=.48\textwidth]{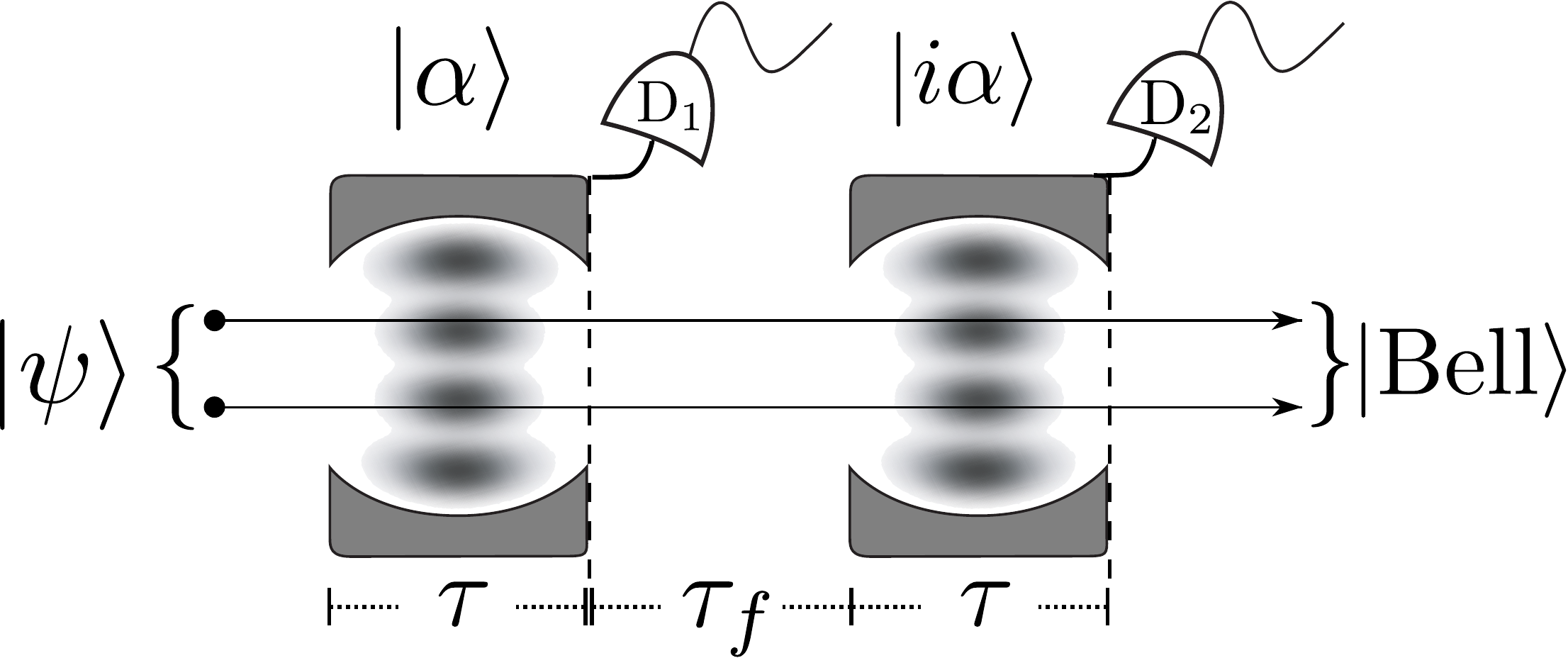}
  \caption{\label{scheme} Schematic representation of the proposed atomic Bell measurement:
  Two atomic qubits interact with the electromagnetic field inside two independent cavities
  in a Ramsey-type interaction sequence. Different projections on the field states 
  inside
  the cavities, recorded by detectors ${\rm D}_1$ and ${\rm D}_2$,
  result in a postselection of atomic Bell states 
  as described in Table \ref{table}.}
\end{figure}

In this section we introduce a protocol which implements
a projection onto four orthogonal atomic 
Bell states of Eq. \eqref{bellstates} for any given initial condition of the
atoms. The scheme we propose requires  interactions between the
atoms with two different cavities as sketched in Fig. \ref{scheme}.
The interaction time between the atoms and the 
electromagnetic field in each cavity is assumed to be
$\tau=1/2$. The field in the first (second) cavity has to be prepared in a 
coherent state $\ket\alpha$ ($\ket{i\alpha}$).
After the interaction with the first cavity the resulting field 
is projected onto the initial state $\ket{\alpha}$. 
In  case of failure a projection onto the state $\ket{-\alpha}$
is performed. 
The projection of the field  postselects the atoms in a state that has 
contribution of only two of the  Bell states. This postselected 
atomic state is taken as initial condition
to interact with a second cavity prepared in the state $\ket{i\alpha}$.
The atoms are assumed to
evolve freely for a time $\tau_f$ before interacting with a second cavity. 
This does not affect the protocol as the free Hamiltonian commutes with the
interaction Hamiltonian in Eq. \eqref{Hamilton}.
After the interaction of the atoms with the second cavity,
the field in the second cavity is projected onto 
$\ket{i\alpha}$ and if this fails another  projection onto the
state $\ket{-i\alpha}$ is performed. 
With this field state projection, the atoms are finally postselected
in a unit fidelity Bell state.
In what follows we discuss in detail all the possible outcomes of the protocol.
There is a finite probability to fail completely 
when none of the coherent state field projections is successful. 
This is discussed in Sec. \ref{Discussion}.

\subsection{Projection onto $\ket{\alpha}$ in the first cavity}
Let us consider a successful projection onto the field state $\ket\alpha$
of the first cavity. In this case  the atoms are postselected in the state
\begin{align}
 \frac{1}{\sqrt{P_1}}
 \left(
  c^-\ket{\Psi^-}+d_{\phi+\pi/2}^+\ket{\Phi^+_{\phi+\pi/2}}
  \right)
  \label{psiat1}
\end{align}
with probability 
$P_1=|c^-|^2+|d_\phi^-|^2$. 
To write this state we have also considered the relations
\begin{align}
  \ket{\Phi_{\phi+\pi/2}^\pm}=-i\ket{\Phi^\mp_{\phi}}, \quad d^\pm_{\phi+\pi/2}=id^\mp_{\phi}.
  \label{pihalf}
\end{align}

The postselected atomic state of Eq. \eqref{psiat1} is taken as initial condition
for the interaction with the second cavity prepared in the coherent state $\ket{i\alpha}$
as depicted in Fig. \ref{scheme}.
Two scenarios are possible for the projection of the field in the 
second cavity. In the first place
we consider a projection onto the coherent state $\ket{i\alpha}$
where the atoms are postselected in the state
$\ket{\Psi^-}$ with probability $P_{11}=|c^-|^2/P_1$. 
This can be verified from Eq. \eqref{Psit} as the new initial state does
not have a contribution of $\ket{\Phi^-_{\phi+\pi/2}}$.
As the projections performed in the first and second cavity are independent events
the state $\ket{\Psi^-}$ can be projected with overall success probability $P_1P_{11}=|c^-|^2$,
the initial probability weight of this state before the protocol.
The second possibility is to project onto the state 
$\ket{-i\alpha}$. In that case
the atoms are postselected in the state
$
\ket{\Phi_{\phi}^-}=i\ket{\Phi_{\phi+3\pi/2}^+}
$ 
provided the condition in Eq. \eqref{ncondition} 
is fulfilled.
This can be verified 
using Eq. \eqref{Psit} with an initial coherent
state $\ket{i\alpha}$
and the atoms initially
in the state of Eq. \eqref{psiat1} that has no contribution of $\ket{\Psi^+}$.
The success probability for this event  is $P_{10}=b|d_\phi^-|^2/P_1$. 
Correspondingly the projection onto the atomic state $\ket{\Phi^-_\phi}$ occurs
with overall success probability $P_1P_{10}=b|d_\phi^-|^2$.
This is proportional to its initial probability weight
but not equal. 
The proportionality factor $b$ is given in Eq. \eqref{factor} and accounts to the imperfect
projection onto the states $\ket{i\alpha_{1/2}^\pm}$.

\begin{table}
\begin{center}
  \begin{tabular}{llll}
    \begin{tabular}{l}
     Field state \\
     in detector \\ ${\rm D}_1$
    \end{tabular}
    & 
    \begin{tabular}{l}
     Field state \\
     in detector \\ ${\rm D}_2$
    \end{tabular}
    & 
    \begin{tabular}{l}
     Atomic \\ state\\
     $\ket{{\rm Bell}}$
    \end{tabular}
    & Probability
    \\
\hline\noalign{\smallskip}
    $\ket{\alpha}$ &$\ket{i\alpha}$ &$\ket{\Psi^-}$ & $|c^-|^2$
    \\
\noalign{\smallskip}
    $\ket{\alpha}$ &$\ket{-i\alpha}$ &$\ket{\Phi^-_\phi}$ & $b|d^-_\phi|^2$
    \\
\noalign{\smallskip}
    $\ket{-\alpha}$ &$\ket{i\alpha}$ &$\ket{\Phi^+_\phi}$ & $b|d^+_\phi|^2$
    \\
\noalign{\smallskip}
    $\ket{-\alpha}$ &$\ket{-i\alpha}$ &$\ket{\Psi^+}$ & $b^2|c^+|^2$
  \end{tabular}
\end{center}
\caption{ 
\label{table}
Summary of the Bell state protocol assisted by photonic state measurements.
The first (second) column indicates the photonic field that has to be 
selected in the first (second) cavity by detector 
${\rm D}_1$ (${\rm D}_2$) in the interaction sequence depicted in
 Fig. \ref{scheme}. The third column indicates the resulting atomic state with
the probability of occurrence given in the last column with 
$b=2/\sqrt{4+\pi^2}\approx0.537$. The protocol fails with probability 
$(1-b)(|d^-_\phi|^2+|d^+_\phi|^2)+(1-b^2)|c^+|^2$.
}
\end{table}

\subsection{Projection onto $\ket{-\alpha}$ in the first cavity}
Now we consider a successful  projection 
onto the coherent state $\ket{-\alpha}$ in the first cavity.
In this situation the atoms are postselected in the state
\begin{align}
  \sqrt{\frac{ b}{P_0}}
  \left(
  c^+\ket{\Psi^+}-
  d^-_{\phi+\pi/2}\ket{\Phi^-_{\phi+\pi/2}}
  \right)
  \label{psiat2}
\end{align}
with probability $P_0=b|c^+|^2+b|d_\phi^+|^2$.
We have used the relations in Eqs. \eqref{property1} and \eqref{pihalf}.

The normalized state of Eq. \eqref{psiat2} is taken as initial condition to interact
with the second cavity prepared in the coherent state $\ket{i\alpha}$.
There are two scenarios in the projection of the second cavity.
First we consider
a successful projection onto the state $\ket{i\alpha}$.
As the initial state of Eq. \eqref{psiat2} 
does not  have any contribution of $\ket{\Psi^-}$, 
the atoms are postselected in the state
$
\ket{\Phi^+_{\phi}}=i\ket{\Phi^-_{\phi+3\pi/2}}
$. 
This occurs with success 
probability $P_{01}=b|d^+_\phi|^2/P_0$.
Thus the state $\ket{\Phi^+_\phi}$ is postselected 
with an overall success probability $P_0P_{01}=b|d^+_\phi|^2$.
A second possible situation is a projection onto the state 
$\ket{-i\alpha}$ in the second cavity.  
In this situation the atoms are postselected in 
the state $\ket{\Psi^+}$.
This can be noted from Eq. \eqref{Psit} 
as the second initial atomic state of Eq. \eqref{psiat2} does not have any contribution
of the state $\ket{\Phi^+_{\phi+\pi/2}}$.
The success
probability of this event is $P_{00}=|c_+|^2b^2/P_0$. It implies an overall success probability
of postselecting state $\ket{\Psi^+}$ of $P_0P_{00}=b^2|c^+|^2$.

\section{Discussion of the protocol}
\label{Discussion}

\begin{figure}
  \includegraphics[width=0.48\textwidth]{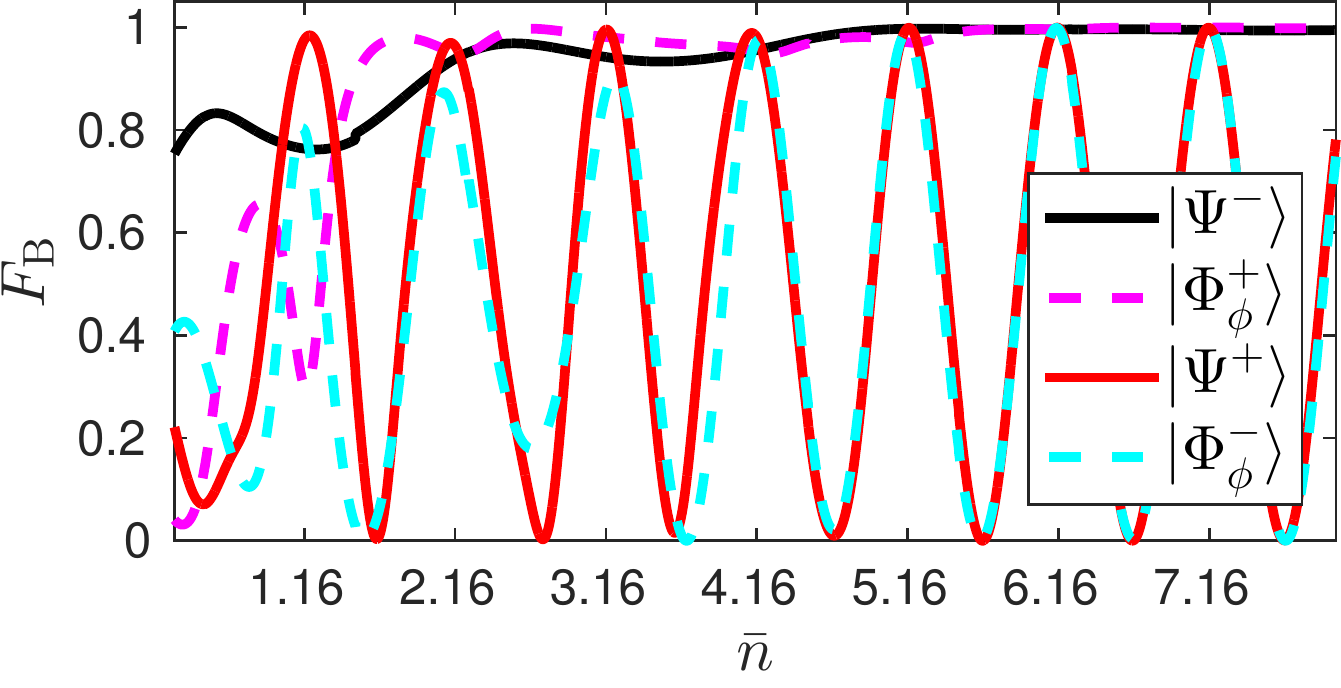}
  \caption{\label{fidelityn}
  Fidelity $F_{\rm B}$ of the projected atomic Bell states as a function of the 
  initial mean photon number of the fields inside the cavities: 
  The interaction time in both cavities is given by 
  $\tau=1/2$, i.e. half  the revival time  (see Eq. \eqref{tau})
  and the rest of the initial conditions are the same as in Fig. \ref{totalfidelity}.
  Each curve correspond to a different Bell state as explained in the legend.
  }
\end{figure}

\subsection{Fidelity of the  postselected Bell states}
In order to test our protocol based on the  approximations
of Eqs. \eqref{Psit} and \eqref{overlaptheta} 
we have numerically evaluated the fidelity 
$F_{\rm B}=|\braket{\rm Bell}{\psi}|^2$ 
of the resulting Bell states in each of the four possible successful outcomes. The state
$\ket{\rm Bell}$ stands for any of the four Bell states of Eq. \eqref{bellstates}.
The state $\ket{\psi}$ is the exact numerical solution after the 
protocol and depends either on the $\bar n$ or $\tau$.
In Fig. \ref{fidelityn} we have plotted the fidelity $F_{\rm B}$
for the different Bell states as a function
of the mean photon number $\bar n=|\alpha|^2$ of the initial coherent 
field states $\ket\alpha$ and $\ket{i\alpha}$.
Interestingly, the protocol already shows high fidelity (above 0.9) even for
small mean photon numbers. The results improve for increasing values of $\bar n$ 
in accordance to the validity of our approximation for high photon number
explained in Sec. \ref{Model}.
The fidelity has an oscillatory periodic 
behaviour and maxima are achieved close to the values of $\bar n$ predicted by Eq. 
\eqref{ncondition}, i.e.
when $\bar n$ is an integer number plus the constant $\arctan(\pi/2)/2\pi\approx 0.16$.
A possible error $\delta  \bar n$ in the previous value has to fulfill 
the condition $\delta \bar n\ll 1/\pi$ to ensure a high fidelity of the atomic states.
It should be mentioned that in the extreme opposite case in which both cavities are 
initially prepared in the vacuum state, i.e. $\bar n=0$, the proposed protocol does not work.
According to Eq. \eqref{Psit} the four orthogonal Bell states are paired up
with three field states and in order to filter out all Bell states they have to be 
orthogonal. This requirement can only be fulfilled in the limit of high mean number of photons.

\begin{figure}
  \includegraphics[width=0.48\textwidth]{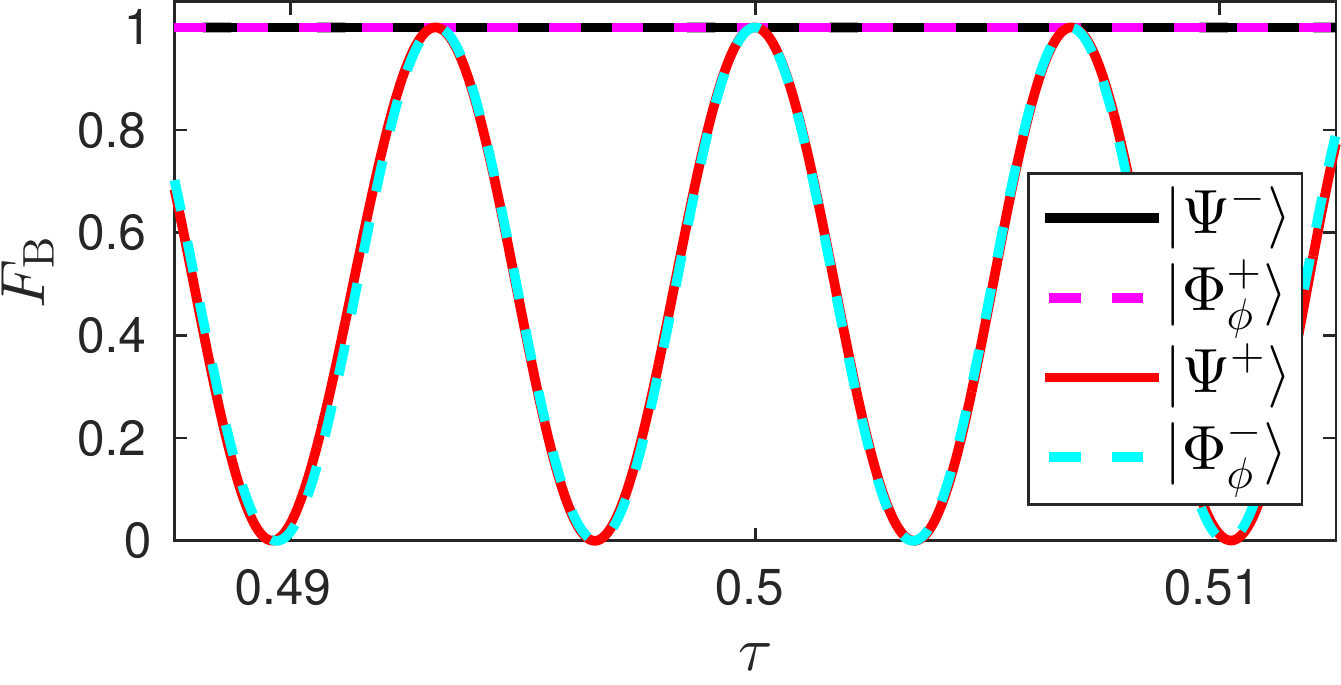}
  \caption{\label{fidelity}
  Fidelity  $F_{\rm B}$ of the projected atomic Bell states as a function of the
  scaled interaction time $\tau$ (see Eq. \eqref{tau}) on both cavities: 
  The initial conditions are the same as in Fig. \ref{wigfig} with $\bar n=36.16$.
  Each curve correspond to a different Bell state of Eq. \ref{bellstates} as 
  explained in the legend.
  }
\end{figure}

To test the sensitivity of the protocol with respect to
the interaction time we have also evaluated the fidelity $F_{\rm B}$ as a 
function of the scaled interaction time
$\tau$ between the atoms and the cavities. 
The results with the initial atomic conditions of Fig. \ref{totalfidelity}
are plotted in Fig. \ref{fidelity}. We present the results for an initial coherent state with 
mean photon number $\bar n= 36+\arctan(\pi/2)/2\pi$. 
The black solid curve represents the fidelity
of projecting onto state $\ket{\Psi^-}$ and it shows a constant unit fidelity
in the time interval of the plot.
The stability of this result has also been discussed in Ref. \cite{Torres2014}
and is due to the fact that $\ket{\Psi^-}$ is a special invariant atomic state of the
two-atom Tavis-Cummings model. 
The fidelity of the state $\ket{\Phi^+_\phi}$  also shows robustness
with respect to the interaction time $\tau$. 
This is due to the fact that this
state is obtained after projecting onto $\ket{i\alpha}$ which is the stationary 
initial state of the second cavity. Advantages of the two-atom Tavis-Cummings model
for generating this particular Bell state have also been mentioned 
previously in Ref. \cite{Rodrigues}.
The other two fidelities of projections onto states
$\ket{\Phi^-_\phi}$ and $\ket{\Psi^+}$
oscillate as a function of $\tau$.  
In this case the second field projection is performed onto field state 
$\ket{-i\alpha}$ and this
in turn has to ``catch'' the time dependent states $\ket{i\alpha_\tau^\pm}$.
Therefore, the oscillations are originated by the overlap between photonic states
$\braket{-\alpha}{\alpha^\pm_{\tau}}$ 
that is calculated  in the Appendix \ref{appendixoverlap}. 
One can estimate that the fidelity $F_{\rm B}$ around $\tau=1/2$ oscillates
with frequency $2(\bar n+1)$. 
The optimal interaction time 
according to Eq. \eqref{overlapepsilon} is $\tau=1/2$ where the
absolute value of the overlap attains its maximum.
A possible error $\varepsilon$ in the 
scaled interaction time $\tau=1/2+\varepsilon$ 
has to be restricted to the condition $|\varepsilon|\ll 1/4\pi(\bar n +1)$.

\subsection{Experimental constraints}
Our protocol requires that the pair of atoms interact with two different coherent states.
This could be realized, for instance, by transporting and positioning
the atomic qubits in separate cavities. Current experimental realizations report coherent transport
and controlled positioning of neutral 
atoms in optical cavities \cite{Reimann,Nussmann2005,Khudaverdyan,Brakhane}, 
where a dipole trap is used as a conveyor belt to displace them. Two trapped ions have also been reported
to be coupled in a controlled way to an optical resonator \cite{Casabone1,Casabone2}. The cavity
can be shifted with respect to ions, allowing to tune the coupling strength between ions and optical cavity. 
In this setting, instead of transporting the atoms to a different cavity, the same cavity might be
shifted to a position where it decouples from the atoms until the measurement is achieved. Then it
would have to be prepared and shifted again for a second interaction with the ions. 

In our discussion we have not considered losses. 
The effects of decoherence can be neglected in the strong coupling regime,
where the coupling strength $g$ between atoms and cavity is much larger than the spontaneous decay
rate of the atoms $\gamma$ and the photon decay rate of the cavity $\kappa$. Actually, in our setting due to the specific interaction time $t_r/2\approx\pi\sqrt{\bar n}/g$ tighter
constraints are required. More specifically, for the cavities we require  
$1/\kappa \gg \pi\sqrt{\bar n}/g$ and for the 
atoms $1/\gamma\gg\pi\sqrt{\bar n}/g$.
The experiment by Khudaverdyan et al. \cite{Khudaverdyan} 
achieved  ratios $g/\kappa=32.5$ and $g/\gamma=5$ which imply  that $ \bar n \ll 2.5$. 
For a single atom interacting with a 
cavity, the experiment by Birnbaum et al. \cite{Birnbaum} involves ratios $g/\kappa=8.26$, $g/\gamma=13.03$ and
if there is a possibility to attain these parameters for a two-atom scenario then the constraint would yield 
$\bar n\ll 7$.
In microwave cavities \cite{Raimond}, the numbers are $g/\kappa\approx 60$ and $g/\gamma\approx 3000$ which
lead to the condition $\bar n\ll 360$.
Thus, the coherence requirement of our proposal is in the reach of current experimental capabilities. 

We have mentioned that our protocol requires the implementation of 
projections onto coherent states. We are not aware of an
experimental solution to this problem. However, coherent states and the vacuum state
are routinely distinguished  in current experiments, see e.g.  \cite{Wittmann}.  
A successful measurement of the vacuum state is achieved when no photons are detected. 
Therefore, for our purposes it would be sufficient to displace the state of the
field in such a way that the field contributions $\ket{\alpha_{1/2}^\pm}$ are
close to the vacuum state. This can be achieved by  driving the optical cavity with a
resonant laser. The Hamiltonian describing this situation in the interaction picture is
$\hat V=\hbar(\Omega^\ast \hat a+\Omega \hat a^\dagger)$.
Under this interaction the states of the field evolve under the influence of 
the evolution operator 
$\hat U_{t_d}=\exp{(-i t_d \hat V/\hbar)}$ that can be identified with the displacement
operator $\hat D(\alpha)=\exp{(\alpha \hat a^\dagger- \alpha^\ast \hat a)}$ provided the
interaction strength of the laser $\Omega$ and the driving time $t_d$ are 
adjusted as $\Omega t_d =i\alpha$. In this way one is able perform the displacement 
$\hat D(\alpha)\ket{-\alpha}=\ket 0$. Finally, we conceive a photodetection 
of the field with three possible outputs; $1)$ No signal, meaning a projection onto
the vacuum state, i.e. $\ket{-\alpha}$ in the undisplaced picture;
$2)$ a weak signal indicating a failure of the protocol; 3)
a strong signal would come from the field state $\ket{2\alpha}$. 

\subsection{Probabilities in the protocol}

A summary of all the possible outcomes of the protocol is given in the Table \ref{table}.
We note that summing the probabilities of all the successful outcomes of the protocol
results in an overall success probability of 
$P_{\rm T}=b+(1-b)(|c^-|^2-b|c^+|^2)$ which depends on the initial state of the system. 
The complementary  probability $1-P_{\rm T}$ corresponds to events that lead to  failure 
of the protocol.
There is a possible failure after a successful projection onto $\ket\alpha$ but unsuccessful
projection onto $\ket{-i\alpha}$. This occurs with probability $(1-b)|d_\phi^-|^2$.
It also might happen
that the projection onto the field state  $\ket{-\alpha}$ in the first cavity is unsuccessful. 
This takes place with probability $(1-b)(|c^+|^2+|d_\phi^+|^2)$. Finally, it is possible that
both projections in the first and second cavity fail with probability $(1-b)b|c^+|^2$.
Summing all these failure probabilities leads to 
$1-P_{\rm T}=(1-b)(|d^-_\phi|^2+|d_\phi^+|^2)+(1-b^2)|c^+|^2$. 

\section{Conclusion}
\label{Conclusions}

We have presented a proposal of an unambiguous 
Bell measurement on two atomic qubits with almost unit-fidelity. 
The theoretical description of the scheme involves the resonant two-atom 
Tavis-Cummings model and a Ramsey-type sequential interaction of both atoms 
with single modes of the electromagnetic field in 
two spatially separated cavities. The first and second cavities 
are initially prepared in coherent 
states $\ket\alpha$ and $\ket{i\alpha}$ respectively. 
The interaction time can be adjusted  by controlling the velocities  of the two atoms
passing trough the cavities. Our discussion has concentrated on basic
properties of the two-atom Tavis-Cummings model in the limit of  high photon numbers. 
We have derived an approximate solution of the dynamical equation which
is expressed as a sum of three terms correlating atomic and field states.
A superposition of two atomic Bell states is correlated with the initial coherent state.
Superpositions of the other two Bell states are correlated
with two time dependent field states.
In phase space these time dependent  contributions of the field state overlap on the 
opposite site to the initial coherent state $\ket\alpha$ ($\ket{i\alpha}$) 
in the first (second) cavity at an interaction time of half the revival time.  
For this reason we have proposed projections onto the two coherent states 
$\ket\alpha$ and $\ket{-\alpha}$ in the first cavity, 
and $\ket{i\alpha}$ and $\ket{-i\alpha}$ in the second cavity.
In order to obtain almost unit fidelity atomic Bell states
the mean photon number has to be restricted to the condition 
given in Eq. \eqref{ncondition}.
Our protocol has a finite error probability due to the imperfect
projection onto the time dependent contributions of the field states 
in the cavities that overlap with $\ket{-\alpha}$ and $\ket{-i\alpha}$.
Nevertheless, the four successful events of our protocol 
summarized in Table \ref{table}  unambiguously project onto four different Bell 
states with almost unit fidelity. 

In  view of current experimental realizations of quantum information protocols in the field 
of cavity quantum electrodynamics the scheme discussed in this work requires cutting edge
technology. An experimental implementation would require accurate control of 
the interaction time and of the average number of photons in the cavity. 
Furthermore, the coherent evolution of the joint system must be preserved. 
This imposes the condition that the characteristic time of photon damping in the cavity
and of atomic decay have to be much larger than the interaction time  that scales with the
square root of the mean photon number in the cavity.
Finally, we point out that the implementation of a von Neumann coherent state projection is, 
up to our knowledge, an open problem that has to be considered in future investigations. 
If these obstacles are overcome, our proposal offers a key component for quantum 
information technology
 such as a multiphoton based hybrid quantum repeater.

\section*{Acknowledgments}
This work is supported by the BMBF project Q.com.

\appendix
\section{Approximations with large mean photon numbers}
\label{appendix}
In this Appendix we present the derivation of the 
time dependent state vector of Eq. \eqref{Psit}. 
It has been shown in  Ref. \cite{Torres2014,Torres2010} 
that the time evolution of any initial state in the form of  Eq. \eqref{initial}
can be obtained from the solution of
the eigenvalue problem of the two-atom Tavis-Cummings Hamiltonian \eqref{Hamilton}.
The exact solution can be written in the following form
\begin{equation}
  \ket{\Psi_t}=
  \ket{0,0}\ket{\chi_t^0}
  +\ket{1,1}\ket{\chi^{1}_t}
  +\ket{\Psi^+}\ket{\chi^+_t} 
  +c_-\ket{\Psi^-}\ket{\alpha}
  \label{psi}
\end{equation}
with the relevant photonic states
\begin{align}
  \ket{\chi^0_t}&=
  c_0\,p_0\ket{0}+
  \sum_{n=1}^\infty 
  \tfrac{
  \sqrt{n}
  \left(
  \xi_{n,t}^-
  -\xi_{n,t}^+\right)
  +\sqrt{n-1}\xi_{n}
  }{\sqrt{2n-1}}
  \ket{n},
  \nonumber\\
  \ket{\chi^1_t}&=\sum_{n=2}^\infty 
  \tfrac{
  \sqrt{n-1}
  \left(
  \xi_{n,t}^-
  -\xi_{n,t}^+\right)
  -\sqrt{n}\xi_{n}
  }{\sqrt{2n-1}}
  \ket{n-2},
  \nonumber\\
  \ket{\chi^+_t}&=
  \sum_{n=1}^\infty 
  \left(
  \xi_{n,t}^-+\xi_{n,t}^+
  \right)
  \ket{n-1},
  \label{fieldstates}
\end{align}
and with the aid of the following abbreviations
\begin{align}
  &\xi_{n,t}^\pm
  =
  \frac{e^{\pm i \omega_n t}}{2}
  \left(
  c_+p_{n-1}\mp
  \tfrac{\sqrt{n}\,c_0 p_n+\sqrt{n-1}\,c_1 p_{n-2}}{\sqrt{2n-1}}
  \right),
  \nonumber\\
  &\xi_{n}
  =
  \frac{\sqrt{n-1}\,c_0 p_n-\sqrt{n}\,c_1 p_{n-2}}{\sqrt{2n-1}},\quad
  \omega_n=g\sqrt{4n-2}.
  \nonumber
\end{align}
The coefficients $p_n$ are initial probability amplitudes of the photon 
number states $\ket n$ of the initial  field state $\ket\alpha$. 
The coefficients $c_0$ and $c_1$ are the initial probability amplitudes 
of the states $\ket{0,0}$ and $\ket{1,1}$ and are related to the probability amplitudes
of the state in Eq. \eqref{initial} by
\begin{align}
  d_\phi^\pm&=
  \frac{c_0 e^{i\phi}\pm c_1 e^{-i\phi}}{\sqrt2}.
  \label{}
\end{align}

The expressions of Eq. \eqref{fieldstates} can be significantly
simplified approximately by taking into account that the field is 
initially prepared in a coherent state $\ket\alpha$ 
with photonic distribution 
$p_n=\exp(-\bar n/2+i\phi)\sqrt{\bar n^{n}/n!}$
and by assuming a large mean photon number $\bar n=|\alpha|^2\gg 1$. 
In such case the photonic distribution has the following
property
\begin{align}
  p_n=\sqrt{\frac{\bar n}{n}}e^{i\phi}p_{n-1}\approx e^{i\phi}p_{n-1}.
  \label{}
\end{align}
Applying this approximation to the states of Eq. \eqref{fieldstates} 
we find the following approximations
\begin{align}
  &\ket{\chi^0_t}\approx
  \sum_{n=1}^\infty 
  \tfrac{
  (c_++d_\phi^+)
  e^{-i\omega_n t}
  -(c_+-d_\phi^+)
  e^{i\omega_n t}
  +2d_\phi^-
  }{2\sqrt2}
  p_{n-1}
  \ket{n},
  \nonumber\\
  &\ket{\chi^1_t}\approx\sum_{n=2}^\infty 
  \tfrac{
  (c_++d_\phi^+)
  e^{-i\omega_n t}
  -(c_+-d_\phi^+)
  e^{i\omega_n t}
  -2d_\phi^-
  }{2\sqrt2}
  p_{n-1}
  \ket{n-2},
  \nonumber\\
  &\ket{\chi^+_t}\approx
  \sum_{n=1}^\infty 
  \tfrac{
  (c_++d_\phi^+)
  e^{-i\omega_n t}
  +(c_+-d_\phi^+)
  e^{i\omega_n t}
  }{2}
  p_{n-1}
  \ket{n-1}.
  \nonumber
\end{align}
In order to simplify these expressions we perform a Taylor expansion in the frequencies
$\omega_n$ around $\bar n+1$ as 
\begin{align}
  \omega_n/g&\approx\sqrt{4 \bar n+2}+2\frac{n-\bar n-1}{\sqrt{4\bar n+2}}
  -2\frac{(n-\bar n-1)^2}{(4\bar n+2)^{3/2}}.
  \label{}
\end{align}
The previous second order expansion is valid provided the third order
contribution multiplied by $gt$ is negligible. This
imposes the  restriction on the interaction time
\begin{equation}
  t\ll \frac{(4\bar n+2)^{5/2}}{4g \bar{n}^{3/2}} \approx \bar n/g.
\end{equation}
For the rescaled time $\tau=gt/\pi\sqrt{4\bar n +2}$ used in the main text 
this implies $\tau\ll\sqrt{\bar n}/2\pi$.
In this approximation the field states can be written as
\begin{align}
  \ket{\chi^0_t}&\approx
  e^{-i\phi}
  \tfrac{
  (c_++d_\phi^+)
  \ket{\alpha_t^{-},-1}
  -(c_+-d_\phi^+)
  \ket{\alpha_t^{+},-1}
  +2d_\phi^-\ket\alpha
  }{2\sqrt2},
  \nonumber\\
  \ket{\chi^1_t}&\approx
  e^{i\phi}
  \tfrac{
  (c_++d_\phi^+)
  \ket{\alpha_t^{-},1}
  -(c_+-d_\phi^+)
  \ket{\alpha_t^{+},1}
  -2d_\phi^-\ket\alpha
  }{2\sqrt2},
  \nonumber\\
  \ket{\chi^+_t}&\approx
  \tfrac{
  (c_++d_\phi^+)
  \ket{\alpha^{-}_t,0}
  +(c_+-d_\phi^+)
  \ket{\alpha^{+}_t,0}
  }{2},
  \label{fieldstates2}
\end{align}
with 
\begin{align}
\ket{\alpha_t^\pm,j}&=
\sum_{n=0}^\infty e^{-\frac{|\alpha|^2}{2}}\frac{\alpha^n}{\sqrt{n!}}
e^{\pm i \left(\nu +2\tfrac{n-\bar n+j}{\nu}-2\frac{(n-\bar n+j)^2}{\nu^3}
\right)g t}
\ket{n}
\nonumber\\
&\approx
e^{\pm ij2\pi\tau}\ket{\alpha^\pm_\tau},
\label{alphas0}
\end{align}
$j\in \{-1,0,1\}$ and $\nu=\sqrt{4\bar n +2}$. 
Furthermore, the states $\ket{\alpha_\tau^\pm}$ are defined by Eq. \eqref{PhotonicStates}.
We neglected the contribution of $j$ in the quadratic term of the 
exponent in  Eq. \eqref{alphas0}.
This can be justified given the fact a Poisson distribution
with high mean value is  almost symmetrically centered around its mean with variance equal to its mean.
This implies that
the maximal relevant value in the quadratic term is given by
\begin{align*}
{\rm max}\left\{  \frac{(n-\bar n+j)^2}{\nu}\right\}\approx
2+\frac{2j}{\sqrt{\bar n}}+\frac{j}{2\bar n},
\end{align*}
which shows that the contribution of $j=-1,0,1$ to this term is negligible for $\bar n\gg 1$.

Finally, using the approximations of Eq. \eqref{fieldstates2} 
and \eqref{alphas0} in Eq. \eqref{psi} and
separating the atomic states accompanying to the photonic states 
$\ket{\alpha_\tau^\pm}$ and $\ket{\alpha}$ yields the result
of Eq. \eqref{Psit}.

\section{Evaluation of 
$\braket{\alpha}{\alpha^\pm_\tau}$
and $\braket{-\alpha}{\alpha^\pm_\tau}$
}
\label{appendixoverlap}
\begin{figure}
  \includegraphics[width=0.48\textwidth]{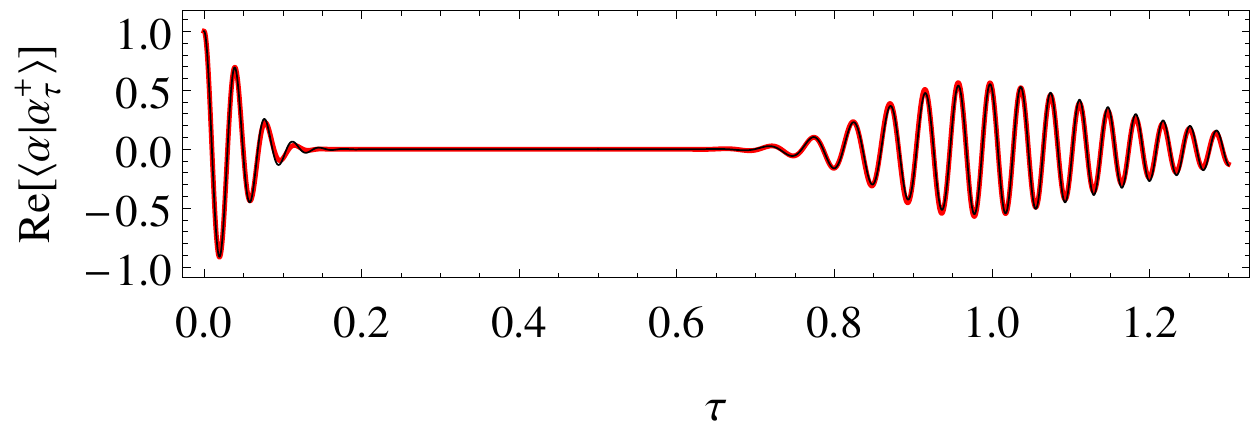}
  \includegraphics[width=0.48\textwidth]{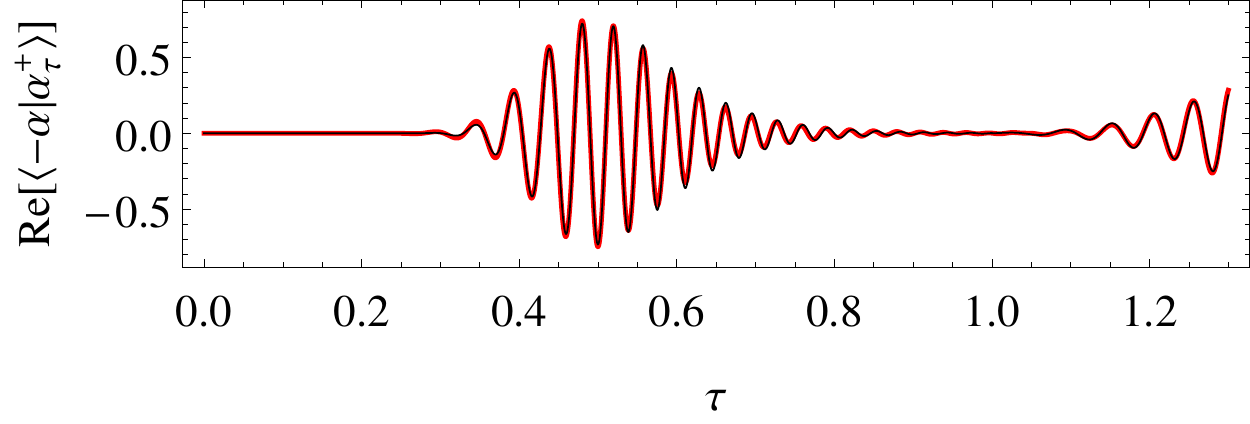}
  \caption{\label{overlaps}
  Top (bottom) figure: Real part  of the overlap 
  $\braket{\alpha}{\alpha^+_\tau}$ 
  ($\braket{-\alpha}{\alpha^+_\tau}$) as function of the rescaled time $\tau$.
  The red curve was evaluated using the exact expression 
  in the first line of Eq. \eqref{overlapapp} and the black narrow line corresponds
  to the approximation given in Eq. \eqref{overlapepsilon}. The mean photon number is   
  $\bar n=12.16$.
  }
\end{figure}
In this appendix we investigate the overlaps between the field states 
$\ket{\alpha^\pm_\tau}$ and $\ket{\pm\alpha}$ 
defined in Eqs. \eqref{PhotonicStates} and \eqref{coherentstate} respectively.
Using the index $j\in\{-1,1\}$ one can write  a single expression for the
four overlaps as
\begin{align}
  &\braket{j\alpha}{\alpha_{\tau}^\pm}=
\sum_{n=0}^{\infty} \frac{\bar n^{n}j^n}{n!e^{\bar n}}
e^{
\pm i 2\pi\tau \left[ \bar n+1+n
-\frac{(n-\bar n)^2}{4\bar n+2}\right]
}
\label{overlapapp}
\\
&\approx
\frac{
e^{\pm i(\bar n +1)2\pi\tau}
}{\sqrt{2\pi\bar n}}
\sum_{n=-\infty}^\infty
e^{
\pm i \pi n\left(2\tau+\frac{1-j}{2}\right)
-\frac{(1\pm i\pi\tau)}{2\bar n}(n-\bar n)^2}
\nonumber\\
&=
\frac{
e^{\pm i(\bar n +1)2\pi\tau}
}{\sqrt{1\pm i\pi\tau}}
\sum_{n=-\infty}^\infty
e^{
\pm i2\pi\bar n\left(\tau+\frac{1-j}{4}\pm n\right)
-\frac{2\pi^2 \bar n}{1\pm i\pi\tau}\left(\tau+\frac{j-1}{4}\pm n\right)^2
}.
\nonumber
\end{align}
In the second line we have approximated the Poisson distribution by a normal distribution
and we have extended the sum to $-\infty$. These 
approximations are valid  in the limit $\bar n\gg 1$.  
In the third line we have used the Poisson summation formula \cite{Bellman} 
which in the case of a Gaussian sum can be expressed as
\begin{align}
  \sum_{n=-\infty}^\infty e^{i2\pi un-s (n-\bar n)^2}=
  \sqrt{\frac{\pi}{s}}\sum_{n=-\infty}^\infty 
  e^{i2\pi \bar n(n+u)
  -\frac{\pi^2}{s}(n+u)^2},
  \nonumber
\end{align}
with ${\rm Re}[s]>0$. 
The last expression   in Eq. \eqref{overlapapp}  
involves a summation of Gaussian terms with variance 
$(1+\pi^2\tau^2)/4\pi^2\bar n$. 
This variance  is very small provided the 
condition  $4\bar n\gg \tau^2$ is fulfilled. If this requirement is met, there exists a dominant 
contribution in the summation  that corresponds
to the value of $n$ where $|\tau+(1-j)/4\pm n|$ achieves its minimum value. This
minimum can be evaluated  as 
\begin{equation*}
  f_j(\tau)=
  {\rm frac}\left(\tau+\tfrac{1-j}{4}+\tfrac{1}{2}\right)-\tfrac{1}{2}, 
\end{equation*}
where ${\rm frac}(x)$ denotes the fractional part of $x$.
By considering only the dominant term of the last summation in Eq.
\eqref{overlapapp} one can find the following approximation of the overlap between
field states
\begin{align}
\braket{j\alpha}{\alpha^\pm_{\tau}}&\approx 
\frac{
e^{\pm i2\pi[\bar n f_j(\tau)+(\bar n +1)\tau]}
}{\sqrt{1\pm i\pi\tau}}
e^{-\frac{2\pi^2 \bar n}{1\pm i\pi\tau}[f_j(\tau)]^2},
\label{overlapepsilon}
\end{align}
with $j\in\{-1,1\}$. 
This result for $\tau=1/2$ and $j=-1$ has been  
rewritten in polar form in Eq. \eqref{overlaptheta} of the main text,
where we used that $f_{-1}(1/2)=0$. In Eq. \eqref{overlapsimple} we have used
that $f_1(1/2)=-1/2$.
In the top panel of Fig. \ref{overlaps} have plotted the real part of the overlap   
$\braket{\alpha}{\alpha^+_\tau}$ as a function of the rescaled time $\tau$.
The evaluation of the exact expression is shown in red and the approximation in black. 
The collapse and revival phenomena are well described by the approximation of the
overlap in Eq. \eqref{overlapepsilon}.
Similar treatment to describe the collapse and revival phenomena in the Jaynes-Cummings
model has been presented in Ref. \cite{Karatsuba}.
In the bottom figure of Fig. \ref{overlaps} we have plotted the real part of  the overlap $\braket{-\alpha}{\alpha^+_\tau}$.


\begin{thebibliography}{99}
\bibitem{Briegel98} H.-J. Briegel, W. D\"ur, J. I. Cirac and P. Zoller,
Phys. Rev. Lett. {\bf 81}, 5932 (1998).
%
\bibitem{Duer99} W. D\"{u}r, H.-J. Briegel, J. I. Cirac and P. Zoller,
Phys. Rev. A {\bf 59}, 169 (1999).
%
\bibitem{Bennett} C. H. Bennett, G. Brassard, S. Popescu,
B. Schumacher, J. A. Smolin, and W. K. Wootters, Phys. Rev.
Lett. {\bf 76}, 722 (1996).
%
\bibitem{Deutsch} D. Deutsch, A. Ekert, R. Jozsa, C.
Macchiavello, S. Popescu, and A. Sanpera, Phys. Rev. Lett.
{\bf 77}, 2818 (1996).
%
\bibitem{Zuk} M. Zukowski, A. Zeilinger, M. A. Horne,
and A. K. Ekert, Phys. Rev. Lett. {\bf 71}, 4287 (1993).
%
\bibitem{Sangouard} N. Sangouard, C. Simon, H. de Riedmatten, and N. Gisin, Rev.
Mod. Phys. {\bf 83}, 33 (2011).
%
\bibitem{vanLoock1} P. van Loock, T. D. Ladd, K. Sanaka, F. Yamaguchi, K. Nemoto, W. J. Munro, and Y. Yamamoto, 
Phys. Rev. Lett. {\bf 96}, 240501 (2006).
%
\bibitem{vanLoock2} T. D. Ladd, P. van Loock, K. Nemoto, W. J. Munro, and Y.
Yamamoto, New J. Phys. {\bf 8}, 184 (2006).
%
\bibitem{vanLoock3} P. van Loock, N. L\"utkenhaus, W. J. Munro, and K. Nemoto, Phys. Rev. A {\bf 78}, 062319 (2008).
%
\bibitem{vanLoock4} D. Gonta and P. van Loock, Phys. Rev. A {\bf 88}, 052308 (2013).
%
\bibitem{Bernad1} J. Z. Bern\'ad and G. Alber, Phys. Rev. A {\bf 87}, 012311 (2013).
%
\bibitem{Bernad2} J. Z. Bern\'ad, H. Frydrych, and G. Alber, J. Phys. B {\bf 46}, 235501 (2013).
%
\bibitem{Torres2014} J.M. Torres, J.Z. Bern\'ad and G. Alber, Phys. Rev. A {\bf 90}, 012304 (2014).
%
\bibitem{Bennett93} C. H. Bennett, G. Brassard, C. Cr\'{e}peau, R. Jozsa, A. Peres, and W. K. Wootters, 
Phys. Rev. Lett. {\bf 70}, 1895 (1993).
%
\bibitem{Bennett92} C. H. Bennett and S. J. Wiesner, Phys. Rev. Lett. {\bf 69}, 2881 (1992).
%
\bibitem{Knill} E. Knill, R. Laflamme and G. Milburn,Nature {\bf 409}, 46 (2001).
%
\bibitem{Pittman} T. B. Pittman, M. J. Fitch, B. C Jacobs, and J. D. Franson, Phys. Rev. A {\bf 68}, 032316 (2003).
%
\bibitem{Munro}  W. J. Munro, K. Nemoto, T. P. Spiller, S. D. Barrett, P. Kok, and R. G. Beausoleil, J. Opt. B: Quantum Semiclass. Opt. {\bf 7}, 135 (2005).
%
\bibitem{Kim}Y.-H. Kim, S. P. Kulik  and Y. Shih,  Phys. Rev. Lett. {\bf 86}, 1370 (2001).
%
\bibitem{Schuck} C. Schuck, G. Huber, C. Kurtseifer, and H. Weinfurter, Phys. Rev. Lett. {\bf 96}, 190501 (2006).
%
\bibitem{Pellizzari} T. Pellizzari, S. A. Gardiner, J. I. Cirac and P. Zoller, 
  Phys. Rev. Lett. {\bf 21}, 3788 (1995).
%
\bibitem{Lloyd} S. Lloyd, M. S. Shahriar, J. H. Shapiro and P. R. Hemmer,
  Phys. Rev. Lett. {\bf 87}, 167903 (2001).
%
\bibitem{Schmidt-Kaler} F. Schmidt-Kaler, H. Häffner, M. Riebe, S. Gulde, G. P. T. Lancaster, T. Deuschle, C. Becher, C. F. Roos, J. Eschner, and R. Blatt,  Nature {\bf 422}, 408 (2003).
%
\bibitem{Isenhower} L. Isenhower, E. Urban, X. L. Zhang, A. T. Gill, T. Henage, T. A. Johnson, T. G. Walker, and M. Saffman, Phys. Rev. Lett. {\bf 104}, 010503 (2010).
%
\bibitem{Noelleke} C. N\"olleke, A. Neuzner, A. Reiserer, C. Hahn, G. Rempe, and S. Ritter, Phys. Rev. Lett. {\bf 110}, 140403 (2013).
%
%
\bibitem{Tavis} M. Tavis and F. W. Cummings, Phys. Rev. {\bf 170}, 279 (1968).
%
\bibitem{Casabone1} B. Casabone, A. Stute, K. Friebe, B. Brandst\"atter, K. Sch\"uppert, R. Blatt, and T. E. Northup, Phys. Rev. Lett. {\bf 111},
100505, (2013).
%
\bibitem{Casabone2} B. Casabone, K. Friebe, B. Brandstätter, K. Schüppert, R. Blatt, and T. E. Northup, Phys. Rev. Lett. {\bf 114}, 023602 (2015).
%
\bibitem{Reimann} R. Reimann, W. Alt, T. Kampschulte, T. Macha, L. Ratschbacher, N. Thau, S. Yoon, and D. Meschede, Phys. Rev. Lett. {\bf 114}, 023601 (2015).
%
\bibitem{Nussmann2005}
  S. Nu\ss{}mann,  M. Hijlkema, B. Weber, F. Rohde, G. Rempe, and A. Kuhn, Phys. Rev. Lett. {\bf 95} 173602 (2005).
%
\bibitem{Glauber} R. J. Glauber, Phys. Rev. {\bf 131}, 2766 (1963).
%
\bibitem{Perelomov} A. Perelomov,   {\it Generalized Coherent States and Their Applications } 
  (Springer-Verlag Berlin Heidelberg 1986).
%
\bibitem{Jarvis} C. E. A. Jarvis, D. A. Rodrigues, B. L. Gy\"orffy, T. P. Spiller, A. J. Short, and J. F. Annett, New J. Phys. {\bf 11}, 103047 (2009).
%
%
\bibitem{Rodrigues} D. A. Rodrigues, C. E. A. Jarvis, B. L. Györffy, T. P. Spiller and J. F. Annett, J. Phys.: Condens. Matter {\bf 20} 075211 (2008).
%
\bibitem{Kim2002} M. S. Kim, J. Lee, D. Ahn, P. L. Knight, Phys. Rev. A, {\bf 65} 040101(R) (2002).
%
\bibitem{Tessier} T. E. Tessier, I. H. Deutsch, A. Delgado, and I. Fuentes-Guridi Phys. Rev. A, {\bf 68} 062316 (2003).
%
\bibitem{Eberly1980} J. H. Eberly, N. B. Narozhny and J. J. Sanchez-Mondragon,  Phys. Rev. Lett. {\bf 44} 1323 (1980).
%
\bibitem{Schleich} W. P. Schleich {\it Quantum Optics in Phase Space} (Wiley-VCH, Weinheim, 2001).
%
\bibitem{Risken} K. Vogel and H. Risken, Phys. Rev. A {\bf 40}, 2847 (1989).
%
\bibitem{Torres2010} J. M. Torres, E. Sadurni, and T. H. Seligman, J. Phys. A {\bf 43}, 192002 (2010).
%
\bibitem{Gea-Banacloche} J. Gea-Banacloche, Phys. Rev. A, {\bf 44} 5913 (1991).
%
\bibitem{Khudaverdyan}M. Khudaverdyan, W. Alt, I. Dotsenko, T. Kampschulte, K. Lenhard, A. Rauschenbeutel, S. Reick, K. Schörner, 
  A. Widera and D. Meschede, New J. Phys., {\bf 10} 073023 (2008).
%
\bibitem{Brakhane} S. Brakhane, W. Alt, T. Kampschulte, M. Martinez-Dorantes, R. Reimann, S. Yoon, A. Widera, and D. Meschede
  Phys. Rev. Lett., {\bf 109} 173601 (2012).
%
\bibitem{Raimond} J. M. Raimond, M. Brune, and S. Haroche, Rev. Mod. Phys., {\bf 73} 565 (2001).
%
\bibitem{Birnbaum} K. M. Birnbaum, A. Boca, R. Miller, A. D. Boozer, T. E. Northup, and H. J. Kimble, Nature {\bf 436}, 87 (2005). 
%
\bibitem{Wittmann} C. Wittmann, M. Takeoka, K. N. Cassemiro, M. Sasaki, Gerd. Leuchs, and U. L. Andersen, 
Phys. Rev. Lett., {\bf 101} 210501 (2008).
%
\bibitem{Bellman} R.E. Bellman, {\it A Brief introduction to theta functions} (Holt, Rinehart and Winston, New York, 1961).
%
\bibitem{Karatsuba}A. A. Karatsuba, and E. A. Karatsuba, J. Phys. A, {\bf 42} 195304 (2009).
%
\end{thebibliography}
\end{document}